\documentclass[aps,prb,twocolumn,floatfix,superscriptaddress,showpacs]{revtex4}
\usepackage{graphicx}
\usepackage{bbm}
\usepackage{amsmath,amssymb}
\newcommand{\be}{\begin{equation}}
\newcommand{\beq}{\begin{equation}}
\newcommand{\ee}{\end{equation}}
\newcommand{\bea}{\begin{eqnarray}}
\newcommand{\eea}{\end{eqnarray}}
\newcommand{\ba}{\begin{array}}
\newcommand{\ea}{\end{array}}
\renewcommand{\vr} {{\bf r}}

\newcommand{\vs} {{\bf s}}

\begin{document}
\title{Local correlation functional for electrons in two dimensions} 

\author{S. Pittalis}
\email[Electronic address:\;]{pittalis@physik.fu-berlin.de}
\affiliation{Institut f{\"u}r Theoretische Physik,
Freie Universit{\"a}t Berlin, Arnimallee 14, D-14195 Berlin, Germany}
\affiliation{European Theoretical Spectroscopy Facility (ETSF)}

\author{E. R{\"a}s{\"a}nen}
\email[Electronic address:\;]{esa@physik.fu-berlin.de}
\affiliation{Institut f{\"u}r Theoretische Physik,
Freie Universit{\"a}t Berlin, Arnimallee 14, D-14195 Berlin, Germany}
\affiliation{European Theoretical Spectroscopy Facility (ETSF)}

\author{M.\,A.\,L. Marques}
\email[Electronic address:\;]{marques@tddft.org}
\affiliation{Laboratoire de Physique de la Mati\`{e}re
Condens\'{e} et Nanostructures, 
Universit\'{e} Lyon I, CNRS, UMR 5586, Domaine scientifique de la
Doua, F-69622 Villeurbanne Cedex, France}
\affiliation{European Theoretical Spectroscopy Facility (ETSF)}

\date{\today}

\begin{abstract}
We derive a local approximation for the correlation energy in
two-dimensional electronic systems. In the derivation we follow the
scheme originally developed by Colle and Salvetti for three
dimensions, and consider a Gaussian approximation for the
pair density.  Then, we  introduce an {\em ad-hoc} modification
which better accounts for  both the long-range correlation,
and the kinetic-energy contribution to the correlation energy.
The resulting functional is {\it local}, and
depends parametrically on the number of electrons in the system. 
We apply this functional to the homogeneous electron gas and
to a set of two-dimensional quantum dots
covering a wide range of electron densities and thus various amounts
of correlation. In all test cases we find an excellent agreement
between our results and the exact correlation energies. 
Our correlation functional has a form that is simple and straightforward
to implement, but broadly outperforms the commonly used local-density
approximation.
\end{abstract}

 
\maketitle

\section{Introduction}

In the last couple of decades, the growing world of nanotechnology put
at our disposal several classes of low-dimensional materials.
Particularly interesting examples are two-dimensional (2D) quantum
dots~\cite{qd1,qd2} (QDs), formed at the interface between two
semiconductors. These systems are not only important from a
technological point of view, but are also remarkable from a purely
theoretical perspective. In fact, as they can be built with different
shapes and sizes, and with a varying number of electrons, they are the
ideal system to study electronic correlation.

The problem of electronic correlation is perhaps the most challenging
in the field of condensed matter physics. Numerous approaches to
handle this problem, with varying degrees of sophistication and
complexity, have been put forward since the very birth of quantum
mechanics. Few-electron QDs can be
studied accurately by, e.g.,
configuration interaction~\cite{rontani} (CI), or by quantum
Monte Carlo techniques~\cite{harju,pederiva,guclu} (QMC). To
describe the electronic properties of larger dots one has to resort to
alternative approaches such as extended Hartree-Fock~\cite{cavaliere} 
(HF) or density-functional theory~\cite{qd2,dft_large,spin_droplet} (DFT).

In DFT, the complexity of the many-body problem is embodied in the
so-called exchange and correlation functional. Several approximations
exist for this quantity, allowing for very accurate calculations of
electronic properties in atoms, molecules, and solids. Clearly, most
exchange-correlation functionals are derived for three-dimensional 
electronic systems.
However, these approximations are known to break down when
applied in the 2D limit.~\cite{kim_pollack} 
This calls for new formulas specialized for
the 2D case. Particularly challenging in these applications is the
fact that, compared with atomic systems, correlation effects in 2D
typically have a more prominent role due to the large size of the
systems (from $10^{-8}$ to $10^{-6}$\,m), and to their low electronic
densities.

Within DFT, 2D systems such as QDs are commonly studied using the 2D
version of the local-density approximation (LDA). It is a combination
of the exchange functional derived for the uniform 2D electron gas by
Rajagopal and Kimball,~\cite{rajagopal} and the corresponding
correlation functional fitted to accurate QMC calculations. The first
of these LDA correlation functionals was put forward by Tanatar and
Ceperley~\cite{tanatar} in 1989. Later on, it was generalized for the
complete range of collinear spin polarizations by Attaccalite {\em et
al.}~\cite{attaccalite} Applications of the 2D-LDA to QDs have been
generally successful, even up to high magnetic
fields.~\cite{qd2,spin_droplet} The LDA, however, suffers from several
shortcomings, already well known from the three-dimensional world, especially for
strongly inhomogeneous systems, or in the low-density (strong
correlation) regime.

Several alternative paths exist to go beyond the simple LDA. A
particularly successful approach starts with the seminal work of Colle
and Salvetti~\cite{CS1,CS2} (CS) who, starting with a physically
motivated ansatz for the many-body wavefunction, developed a closed
formula for the correlation energy. This formula has received a large
interest, especially because it was used to derive the popular
Lee-Yang-Parr (LYP) generalized gradient
functional:~\cite{LeeYangParr:88} Together with Becke's exchange
functional~\cite{Becke:88} it forms the BLYP functional, and in hybrid
schemes it is a part of B3LYP,~\cite{Becke:93} X3LYP,~\cite{x3lyp}
etc.

Interestingly, the same CS formula can also be interpreted as an
orbital-dependent correlation functional, especially suited for DFT
calculations beyond the exact-exchange.~\cite{GraboGross:95_Pittalis2} It
should, however, be emphasized that the CS correlation-energy
functional has several known
limitations.~\cite{SinghMassaSahni:99,TaoGoriPerdewMcWeeny:01,ImamuraScuseria:02}
In particular, while short-range correlations are well
described,~\cite{TaoGoriPerdewMcWeeny:01} important long-range
correlations are missing. Even if these latter effects often cannot be
ignored in large molecules and solids, they can be energetically
negligible in small systems such as atoms.  However, 
it has been  shown recently that the long-range correlation problem
may be cured to some extent. \cite{MFA,M} 
Secondly, in the CS
functional the kinetic-energy contribution to the correlation energy
 (named below as the kinetic-energy correlation) is taken 
into account only in an empirical fashion through the fitting parameter.
In this context, an interesting modifications of the original 
CS approach have been recently proposed.~\cite{ragot}

In this work, we generalize the CS scheme~\cite{CS1,CS2} to 2D. 
Then we use a Gaussian approximation for the pair probability
function. Finally, 
we introduce an {\em ad-hoc} modification which, {\it post-factum}, 
seems to recover both the long-range and the kinetic-energy
correlation to some good extent.

\section{Theory}

Our starting point is
the following ansatz~\cite{CS1,CS2} for the many-body wavefunction
$\Psi$
\begin{multline}\label{cs}
  \Psi(\vr_1 \sigma_1,..., \vr_N \sigma_N) =
  \Psi_{\rm SD}(\vr_1 \sigma_1,..., \vr_N \sigma_N) \\
  \times \prod_{i < j} \left[1-\varphi(\vr_i,\vr_j) \right]
  \,.
\end{multline}
Here, $\vr$ and $\sigma$ denote respectively the space and spin
coordinates of the electrons, and $\Psi_{\rm SD}$ indicates the 
{\em single} Slater determinant of HF theory, which in the DFT context
should be replaced by the Slater determinant generated from the occupied
Kohn-Sham orbitals. The function $\varphi$ describes the correlated
part of the wavefunction. In the center-of-mass,
$\vr=(\vr_1+\vr_2)/2$, and relative, $\vs=\vr_1-\vr_2$, coordinate
system, it can be written as
\be\label{cf}
  \varphi(\vr,s) = \left[ 1 - \Phi(\vr)(1 + \alpha s) \right]e^{-\beta^2(\vr) s^2}
  \,,
\ee
where the quantities $\Phi$, $\alpha$, and $\beta$ act as correlation
factors. We point out that we introduce $\beta(\vr)$ as a {\em local}
$\vr$-dependent quantity for reasons which become obvious below. To
find a reasonable value for $\beta$, which determines the local correlation
length, we estimate the area where the electron is
correlated as
\be
  A(\vr) = \int\!\!d^2s\; e^{-\beta^2(\vr) s^2} = \frac{\pi}{\beta^2(\vr)}
  \,.
\ee
Then, we assume that this area is proportional to the area of 
the Wigner circle $\pi r_s^2$, where the density parameter $r_s$ 
is given through the total electron density as $r_s(\vr)=1/\sqrt{\pi\rho(\vr)}$.
Thus, we find the relation
\be\label{beta}
  \beta(\vr)=q\sqrt{\rho(\vr)}
  \,,
\ee
where $q$ is a fitting parameter. 

The SD wavefunction in Eq.~\eqref{cs} is recovered when all pairs of
electrons are far apart from each other. In contrast, when two
electrons are brought to the same point, the parameter $\alpha$ is
chosen to satisfy the {\em cusp condition} (for the
singlet case) of the wavefunction.  It can be shown~\cite{RJK} that,
for the 2D case, $\alpha=1$.  The function controlling the exponential
decay is given by
\be\label{Phi}
  \Phi(\vr) = \frac{\beta(\vr)}{\beta(\vr) + \sqrt{\pi}/2},
\ee
which can be deduced by imposing the condition~\cite{CS2,MS}
\be
  \int\!\!d^2s\; \varphi(\vr,\vs) = 0
  \,.
\ee

By using the wavefunction~\eqref{cs} and the definition of the
correlation factor $\varphi$ given by~\eqref{cf}, we can obtain a
formula for the correlation energy~\cite{CS1,CS2}
\be\label{ce2}
  E_c = \int\!\!d^2r\!\!\int\!\!d^2s\;
  \rho_{2,{\rm SD}}(\vr,\vs) \frac{\varphi^2(\vr,\vs)-2\varphi(\vr,\vs)}{s}
  \,,
\ee
where $\rho_{2,{\rm SD}}(\vr,\vs)$ refers to the SD pair density. To
simplify this expression, we write a Gaussian approximation for this
function,
\be
  \rho_{2,{\rm SD}}(\vr,\vs) = \rho_{2,{\rm SD}}(\vr) 
  e^{-s^2/\gamma^2(\vr)}
  \,.
\ee
The use of this Gaussian approximation was proposed in the context of
the CS scheme by Moscard\'o and San-Fab\'ian,~\cite{MS} but it has
been used in the field of DFT even further back.~\cite{Parr88} To
obtain the function $\gamma(\vr)$, that defines the width of the
Gaussian, we apply the exact sum-rule
\begin{align}
  \rho_{{\rm SD}}(\vr) & =  \frac{2}{N-1}\int\!\!d^2s\;\rho_{2,{\rm SD}}(\vr,\vs)\nonumber\\
  & =  \frac{2\pi}{N-1}\,\rho_{2,{\rm SD}}(\vr)\gamma^2(\vr)
  \,,
\end{align}
from which follows
\be
  \gamma(\vr) = \sqrt{\frac{(N-1)\,\rho_{{\rm SD}}(\vr)}{2\pi\,\rho_{2,{\rm SD}}(\vr)}}
  \,.
\ee 
To simplify this expression, we apply the relation
\be\label{hfrel}
  \rho_{2,{\rm SD}}(\vr)=\frac{1}{4}\rho_{{\rm SD}}^2(\vr)
  \,,
\ee
as well as Eq.~\eqref{beta} for the SD density $\rho_{{\rm SD}}(\vr)$, 
and find
\be
  \frac{1}{\gamma^{2}(\vr)}=c\,\beta^2(\vr)
  \,,
\ee
where
\be\label{qc}
  c = \frac{\pi}{2(N-1)q^2}
  \,.
\ee
Using these results in Eq.~\eqref{ce2}, and performing the integration
over $s$, leads to the final result
\be\label{Ec}
  E_{\rm c}^{\rm local} = \int\!\!d^2r\; \rho_{{\rm SD}}(\vr)\,\epsilon^{\rm local}_{\rm c}(\vr)
  \,,
\ee
where we have defined $\epsilon_{\rm c}(\vr)$ as the local correlation energy 
per electron having the lengthy expression
\begin{multline}\label{epsilon}
  \epsilon_{\rm c}(\vr) = \frac{\pi}{2q^2}\Bigg\{
    \frac{\sqrt{\pi}\:\beta(\vr)}{2\sqrt{2+c}}[\Phi(\vr) - 1]^2
   +\frac{\Phi(\vr)[\Phi(\vr)-1]}{2+c}
  \\   + \frac{\sqrt{\pi}\:\Phi^2(\vr)}{4\beta(\vr)(2+c)^{3/2}}
   + \frac{\sqrt{\pi}\:\beta(\vr)}{\sqrt{1+c}}[\Phi(\vr)-1]
   + \frac{\Phi(\vr)}{1+c}
  \Bigg\}\,.
\end{multline}

Up to this point, the only inputs for the correlation energy are the fitting parameter
$q$ (we will come back to the choice of this parameter later on), 
the total number of electrons $N$, and the electron density
$\rho(\vr)$. We remind that the parameter $c$ is defined through $q$
and $N$ in Eq.~\eqref{qc}, and $\beta(\vr)$ is given in terms of
$\rho(\vr)$ in Eq.~\eqref{beta}.  
This particular depencency on $N$ 
conflicts with the extent requirement of the correlation
functional. For example, situations where two systems are very far 
apart from each other are expected to be problematic.

In conclusion, equation~\eqref{Ec}
is an {\em explicit density functional} for the correlation energy
with a single fitting parameter $q$.  This functional is
self-interaction free, in the sense that it is identically zero for 
one-electron systems. Note that to recover this important property within
the standard ladder of exchange-correlation functionals, one has to
resort to highly sophisticated orbital functionals.

\section{Application and refinement of the approximation}

Here we complete and apply the approximation for the correlation energy in 2D.
In particular, along the applications, we shall present
an {\em ad-hoc} modification which better accounts for both the
long-range and the kinetic-energy correlation.

 As the first step, now we need to choose a value for the fitting parameter $q$, we use Taut's
analytic result~\cite{taut} for the singlet state of a two-electron
parabolic QD with confining strength $\omega=1$.  In terms of energy
components, the correlation energy can be written as $E_{\rm c} =
E_{\rm tot}-E^{\rm EXX}_{\rm tot}$, where EXX refers to the exact-exchange
result.  Applying this formula yields $E_c\approx
-0.1619$ for the $N=2$ singlet when $\omega=1$.  To obtain the same
value from Eq.~\eqref{Ec}, we need to set $q=2.258$.
Of course the  
choice may be refined, if needed. But aiming at providing 
ideally a {\em predictive} approximation, the fitting
should not be carried out 
for each new system (to 
obtain every time the correct answer) but 
rather carried out once for ever. 
This is a quite general, and a well known way of defining, or refining, 
new approximations for the central quantities of DFT. 
In the following, we will show that our fitting procedure, outlined just above, guarantees 
a very good performance for a large class of systems.

\begin{table}[t]
  \caption{\label{table_qd} Comparison of the correlation energies (in
  atomic units) for parabolic quantum dots. The reference value $E_c^{\rm ref}$
  is obtained by subtracting the exact-exchange energy from accurate
  data for the total energy. The last row contains the
  mean percentual error, $\Delta$, for the parabolic dots (excluding the one
  used in the fitting procedure).
}
  \begin{tabular}{c c c c c c c c}
  \hline
  \hline
  $N$ & $\omega$ & $E_{\rm tot}^{\rm ref}$ & $E^{\rm EXX}_{\rm tot}$ & 
  $-E_{\rm c}^{\rm ref}$ & $-E_{\rm c}^{\rm local}$ & 
  $-E_{\rm c,mod}^{\rm local}$ & $-E_{\rm c}^{\rm LDA}$  \\
  \hline
  2  & 1           & $3^\dagger$               &  3.1619 & 0.1619 & 0.1619$^*$ & 0.1619$^*$ & 0.1988 \\
  2  & 1/4         & $0.9324^\ddagger$         &  1.0463 & 0.1137 & 0.0957 & 0.1212 & 0.1391 \\
  2  & 1/16        & $0.3031^\ddagger$         &  0.3732 & 0.0701 & 0.0477 & 0.0757 & 0.0852 \\
  2  & 1/36        & $0.1607^\ddagger$         &  0.2094 & 0.0487 & 0.0299 & 0.0527 & 0.0607 \\
  6  & 0.42168     & $10.37^{\mathchar "278}$  & 10.8204 & 0.4504 & 0.3805 & 0.4453 & 0.5305 \\
  6  & $1/1.89^2$  & $7.6001^{\mathchar "27B}$ &  8.0211 & 0.4210 & 0.3205 & 0.4060 & 0.4732 \\
  6  & 1/4         & $6.995^\ddagger$          &  7.3911 & 0.3961 & 0.3047 & 0.3946 & 0.4574 \\
  12 & $1/1.89^2$  & $25.636^{\mathchar "27B}$ & 26.5528 & 0.9168 & 0.6837 & 0.8504 & 1.0000 \\
  \hline
  \multicolumn{5}{l}{$\Delta$}                                    & 26.1\% & 5.9\%  & 18.4\% \\
  \hline
  \hline
  \end{tabular}
  \begin{flushleft}
  $^*$         Fitted result (see text).
  $^\dagger$   Analytic solution by Taut from Ref.~\onlinecite{taut}.
  $^\ddagger$  CI data from Ref.~\onlinecite{rontani}.
  $^{\mathchar "278}$ Variational QMC data from Ref.~\onlinecite{lda}.
  $^{\mathchar "27B}$ Diffusion QMC data from Ref.~\onlinecite{pederiva}.
  \end{flushleft} 
\end{table}

\begin{table}
  \caption{\label{table_qd_recta} Comparison of the correlation energies (in
  atomic units) for square ($\pi\times\pi$) quantum dots. The reference value $E_c^{\rm ref}$
  is obtained by subtracting the exact-exchange energy from the
  quantum Monte Carlo result for the total energy (Ref.~\onlinecite{rectapaper}). 
}
  \begin{tabular}{c c c c c c c }
  \hline
  \hline
  $N$ & $E_{\rm tot}^{\rm QMC}$ & $E^{\rm EXX}_{\rm tot}$ & -$E_{\rm
  c}^{\rm ref}$ & -$E_{\rm c}^{\rm local}$ & -$E_{\rm
  c,mod}^{\rm local}$ & -$E_{\rm c}^{\rm LDA}$ \\
\hline
   2 &   3.2731 &   3.4643 & 0.1908 & 0.1905 & 0.1763 & 0.2226 \\
   6 &  26.9679 &  27.5928 & 0.6249 & 0.6578 & 0.5763 & 0.7624 \\
   8 &  46.7940 &  47.5962 & 0.8022 & 0.9168 & 0.7836 & 1.0514 \\
  12 & 103.3378 & 104.5620 & 1.2242 & 1.4494 & 1.2026 & 1.6419 \\
  16 & 178.5034 & 179.9804 & 1.4770 & 2.0096 & 1.6282 & 2.2534 \\
  \hline
  \multicolumn{4}{l}{$\Delta$}      & 19.3\% & 6.8\%  & 33.6\% \\
  \hline
  \hline
  \end{tabular}
\end{table}

Tables~\ref{table_qd} and \ref{table_qd_recta} show results for
parabolically confined, and for square ($\pi\times\pi$) quantum
dots. The results obtained with our local formula for the correlation
energy (denoted by $E_{\rm c}^{\rm local}$) are compared to reference
results $E^{\rm ref}_{\rm c}$, as well as with LDA correlation
energies $E^{\rm LDA}_{\rm c}$.  We computed the EXX and LDA values
using the real-space code {\tt octopus}.~\cite{octopus} 
The EXX result was calculated in the Krieger-Li-Iafrate 
(KLI) approach~\cite{KLI}, which is an accurate
approximation in the static case.~\cite{kli_oep} For $E_{\rm
c}^{\rm LDA}$ we applied the parametrization of Attaccalite {\em et
al}.~\cite{attaccalite} Note that we used a perturbative approach to
calculate $E_c^{\rm local}$ from Eq.~\eqref{Ec}: The self-consistent
EXX density was the input for our local functional. We also found that
using the LDA density as input did not make a considerable difference.

The QDs studied here span a wide range of density parameters $r_s$
determined in a parabolic QD as $r_s=N^{-1/6}\omega^{-2/3}$ and
in our square QD as $r_s=\sqrt{\pi/N}$.
This parameter corresponds to the average
radius of an electron in a QD with an average number density
$n_0=1/(\pi r_s^2)$. Thus, the cases shown in the tables are between
$0.44<r_s<9.71$. In fact, the upper limit exceeds the threshold of
$r_s\sim 7.5$ for Wigner crystallization in the impurity-containing 2D
electron gas.~\cite{chui} One should, however, bear in mind that in
QDs the concept of Wigner localization is ambiguous, and no general
formula exists for the density parameter at the onset of
localization. It can be also seen that in our examples the ratio of
the correlation to the total energy varies from less than one percent
up to around $30\%$.

\begin{figure}
\setlength{\unitlength}{0.95\columnwidth}
\begin{center}
  \includegraphics[width=\unitlength]{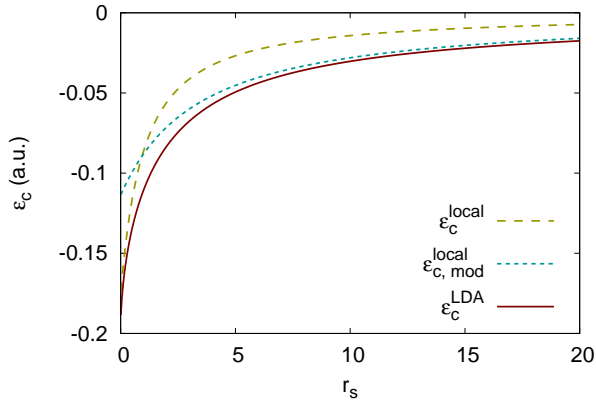}
  \vspace{-1cm}
\end{center}
\caption{\label{fig:egas}
Correlation energy per unit particle for the uniform 2D electron gas in various approximations.
}
\end{figure}

Results with our local formula are roughly of the same quality as the
LDA, slightly worse for the parabolic dots, but slightly better for
the square dots. Furthermore, a calculation for the homogeneous
electron gas (Fig.~\ref{fig:egas}) 
reveals that this functional agrees
with the LDA (which is exact for this system) in the limit of
vanishing $r_s$, but underestimates the correlation energy otherwise.

The derived functional not only gives already very reasonable results, but is
also a very good starting point for further developments. In fact, we
found an alternative functional (that we will denote by $E_{\rm
c,mod}^{\rm local}$) obtained by modifying the first term between the
parenthesis in Eq.~(\ref{epsilon}) by
\be
  [\Phi(\vr) - 1]^2 \rightarrow
  \Phi(\vr) - 1\,.
\label{mod_formula}
\ee
To finish the derivation of this new functional we need to refit the
parameter $q$, which now reads $q^{\rm mod}=3.9274$ (in such a way,
we again obtain the exact value of the correlation energy for the
two-electron quantum dot as described above). 

{\it Sensu stricto}, Eq.~(\ref{mod_formula}) is an empirical approximation.
However, our results suggest, {\it Post-factum}, that the proposed modification 
better accounts for the long-rage~\cite{TaoGoriPerdewMcWeeny:01,M} and
kinetic-energy correlation.~\cite{ragot,umrigar}

According to Tables~\ref{table_qd} and \ref{table_qd_recta}, our
corrected functional agrees very well with the reference results. We
find that, in all the cases studied, our approximation is vastly
superior to the LDA correlation. Note that our results exhibit the
correct scaling with respect to both confinement strength and number
of electrons, even if the adjustable parameter $q$ has only been
fitted to the case $N=1$ and $\omega=2$. Also for the homogeneous
electron gas (Fig.~\ref{fig:egas}), 
our modified functional yields results that are
remarkably close to the reference LDA curve, departing significantly
from the exact curve only for very small $r_s$ (weak correlation
limit).

Finally, we wish to make a few remarks on the usage of the present correlation
functional. 
First, we point out that in practical purposes within, e.g., the Kohn-Sham
scheme of DFT, the functional should be
combined with an adequate recipe for the exchange energy, such as the
exact-exchange or the functionals suggested in
Ref.~\onlinecite{own_Ex}. 
Second, for many systems ---like, e.g.,
QDs in magnetic fields--- one requires a spin-polarized
version of the exchange-correlation functional. This has already been
taken into account in the LDA functional by Attaccalite
{\em et al.}~\cite{attaccalite}, but a spin-polarized extension of the
present
functional is still missing. Work to solve these two issues is already
under way.

\section{Conclusions}

We developed a correlation energy functional for the
two-dimensional electron gas, starting from the Colle and Savetti
ansatz for the many-body wavefunction and a Gaussian approximation to
the pair density. To better account for the long-range and
  kinetic-energy correlation, we have then introduced an additional 
ad-hoc modification. The resulting functional has a very simple form, depending
parametrically on the total number of electrons $N$ and locally on the
electronic density $n(\vr)$. It only contains a single parameter, $q$
that was adjusted to the exact calculation of a two-electron quantum
dot. Calculations performed for several systems, with a wide range of
density parameters $r_s$, show that our functional gives results in
very good agreement with reference values. This agreement is
maintained even for very dilute electron gases, where the correlation
energy amounts to 30\% of the total energy. Furthermore, our
functional performs significantly better than the standard LDA
correlation functional, while
maintaining much of its simplicity.

\acknowledgments
We thank Ari Harju for the original quantum Monte Carlo data
shown in Table~\ref{table_qd_recta}.
This work was supported by the EU's Sixth Framework Programme through
the Nanoquanta Network of Excellence (NMP4-CT-2004-500198),
Deutsche Forschungsgemeinschaft, and the Academy of Finland.  
M.A.L. Marques
acknowledges partial support by the Portuguese FCT through the project
PTDC/FIS/73578/2006.

\end{document}